\documentclass{emulateapj}

\newcommand{\msun}{M$_{\odot}$}

\shorttitle{NGC\,1313: Evidence for Infant Mortality}
\shortauthors{Pellerin et al.}

\begin{document}


\title{Stellar Clusters in NGC 1313: Evidence for Infant Mortality}


\author{Anne Pellerin \& Martin Meyer}
\affil{Space Telescope Science Institute, 3700 San Martin Drive, Baltimore, MD 21218; \\ pellerin@stsci.edu, martinm@stsci.edu}
\author{Jason Harris}
\affil{Stewart Observatory, 933 North Cherry Avenue, Tucson, AZ 85721; jharris@as.arizona.edu}

\and

\author{Daniela Calzetti}
\affil{University of Massachusetts, Department of Astronomy, 710 North Pleasant Street, Amherst, MA 01003, and \\
Space Telescope Science Institute, 3700 San Martin Drive, Baltimore, MD 21211; calzetti@astro.umass.edu}

\begin{abstract}
We present evidence that infant mortality of stellar clusters is likely to be a major and very efficient process for the dissolution of young clusters in the spiral galaxy NGC\,1313. Performing stellar PSF photometry on archival HST/ACS images of the galaxy, we find that a large fraction of early B-type stars are seen outside of star clusters and well spread within the galactic disk, consistent with the scenario of infant mortality. We also calculate the UV flux produced by the stars in and out the clusters and find that 75 to 90\% of the UV flux in NGC\,1313 is produced by stars outside the clusters. 
These results suggest that the infant mortality of star clusters is probably the underlying cause of the diffuse UV emission in starburst galaxies. Infant mortality would also explain the numerous B-type stars observed in the background field of our Galaxy as well. We exclude the possibility that unresolved low-mass star clusters and scaled OB associations might be the main source for the diffuse UV emission.

\end{abstract}


\section{Introduction}
\label{intro}

Stellar clusters are one of the fundamental components in galaxy formation and evolution since they are the nursery of most stars \citep{lada03}. Given this, one might expect that most UV flux, essentially produced by massive OB stars, would come from star clusters. However, \citet{meurer95} found that around 80\% of the UV flux at 2200\AA\ was produced outside star clusters and from a diffuse source. Similar results were found in galaxies with circumnuclear star-forming rings \citep{maoz96} and normal spiral galaxies \citep{hoopes01}. This fraction does not seem consistent with the idea that clusters are the nursery of stars and several scenarios have been built up to explain these observations.

One possible source for the diffuse UV emission in galaxies are UV photons from clusters that have
been scattered by dust in the 
inter-cluster environment \citep[e.g.][]{pop05}.  However \citet{tremonti01} and \citet{chandar03,chandar05} found that the spectral UV lines from clusters are different from the inter-cluster environment, ruling out this possibility. They clearly showed, using evolutionary spectral synthesis models, that the UV stellar line profiles of star clusters in starburst galaxies is dominated by O-type stars while the inter-cluster environment is dominated by less massive B-type stars. The presence of an underlying B-type star population was also suspected by \citet{hoopes01} as the origin of the diffuse UV emission component in normal spiral galaxies.
These works revealed that scattered photons cannot be the main source of diffuse UV emission, and that a stellar origin must contribute a significant fraction of the UV flux.

A second possible explanation is that unresolved lower-mass clusters or scaled OB associations 
\citep[SOBAs;][]{maiz04} are present between the detected compact clusters \citep{barth95,chandar05}. This is consistent with the general idea first proposed by \citet{meurer95} that there exist two modes of star formation: prominent cluster formation and dominant diffusely distributed star formation. In this scenario, the diffusely distributed star clusters are small enough so their massive stars produced are spectral type B. It also implies that there is ongoing star-formation outside clusters. 

A third possibility is that the massive B stars observed in the diffuse field were formed in clusters that have since dissolved. The only scenario capable of dissolving a large fraction of star clusters over a short period of time is the infant mortality of stellar clusters \citep{lada03,fall05,bast06}. The main idea is that most clusters do not survive the loss of gas and dust expelled during the evolution and death of the most massive stars, the remaining unbound cluster then rapidly dissolves into the diffuse field. This scenario differs from the second one by the fact that the stars are not closely related to their birthplace, avoiding the need for ongoing star-formation activity in the inter-cluster region.

In this letter, we discuss the possibility that massive stars from dissolved clusters are the origin of the diffuse UV emission first observed by \citet{meurer95} and the source of the stellar population observed by \citet{chandar03,chandar05} in inter-cluster regions. We also present indications that the phenomenon of infant mortality occurs in the normal spiral galaxy NGC\,1313.

\section{PSF Photometry of HST/ACS Images}
\label{data}

\begin{figure}
\plottwo{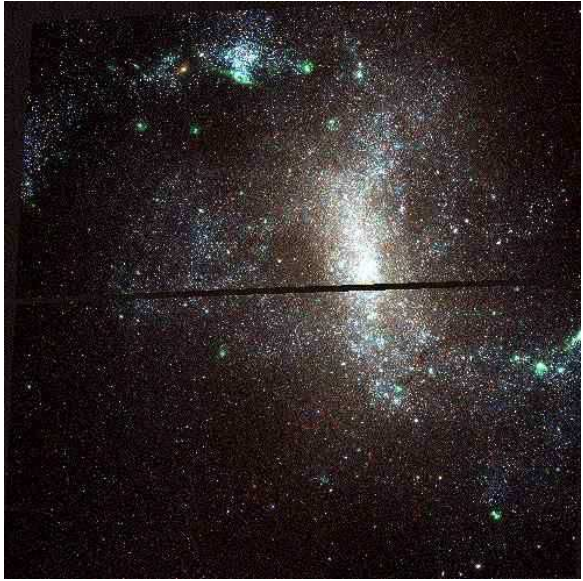}{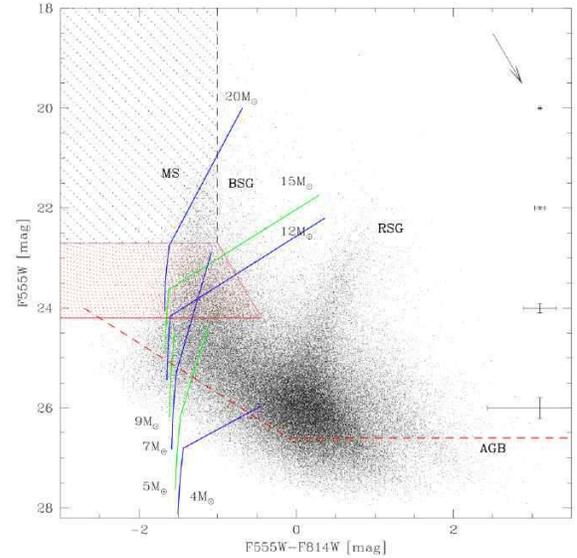}
\caption{\label{cmd} Left: False color ACS/HST image of the central region of NGC\,1313 with blue for F435W, green for F555W, and red for F814W. North is up and east on the left. Right: Color-magnitude diagram of resolved stars in NGC\,1313 for the F555W and F814W filters. Stellar evolutionary tracks for ACS filters from the Padova group are identified in blue and green for various masses. Typical photometric uncertainties are displays on the right. The reddening vector of 1 mag is shown on the upper right corner. The dashed red line represents the 50\% detection limit. The main sequence (MS), blue supergiant (BSG), red supergiant (RSG), and asymptotic giant (AGB) branches are also identified.}
\end{figure}

NGC\,1313 is a face-on SB(s)d galaxy at a distance of $\sim$4.2\,Mpc \citep{tully88,men02,kar04} and shows several star clusters and complexes (see Fig.~\ref{cmd}). We use archival images of NGC\,1313 obtained with the Advance Camera for Surveys (ACS) onboard the Hubble Space Telescope (HST) for Cycle~12 project ID9796. For this galaxy, the spatial scale is 1.0\,pc pixel$^{-1}$ (0.05\,arcsec pixel$^{-1}$) and the spatial resolution is about 2\,pc, which allows us to resolve individual bright stars. The ACS field covers the central bar, most of the northern spiral arm, and about half of the southern spiral arm. Images in the three wide-band filters F435W, F555W, and F814W are available for the field-of-view centered on the nucleus.  The data were reduced by the STScI calibration pipeline, which includes treatment for cosmic rays and drizzling\footnote{\url{http://.stsci.edu/hst/acs/analysis/}}. 

We performed PSF photometry on the F435W, F555W and F814W images using the IRAF/DAOPHOT package. PSF photometry is required because of crowding. We used about 100 bright unsaturated stars to create a PSF model with the form of a Moffat function with 2-3 pixels FWHM having a 15 pixels radius. Note that the PSF varies within an image, which contributes to the photometric uncertainties. The absolute photometric calibration is calculated from the zero points given by \citet{sirianni05}. We calculate an aperture correction of -0.229, -0.192 and -0.27 mag for the F435W, F555W, and F814W filters respectively. The 50\% detection limits are reached at 26.0, 26.6 and 26.7 magnitudes for the F435W, F555W, and F814W filters respectively. Because it is not possible to correct for the extinction of individual stars, we simply apply a global correction of E(B$-$V)=0.1 to each image \citep{helm04},
which corresponds to the foreground extinction. We will discuss the impact of a larger uncertainty value later in the text. Finally, we compared the total flux of the images to the integrated flux from the PSF stars and find that about 1\% of the total flux in the three filters is coming from diffuse unresolved and/or non-stellar sources. Considering the spatial resolution of 2\,pc, it is possible that some sources are not individual stars but compact low-mass clusters, similar to Galactic open clusters \citep[e.g.][]{janes88,bica03,schi06}, having a very few or no B-type stars. This point will be discussed later on.

\section{Color-Magnitude Diagram and Spatial Maps}
\label{analy}

We produced a color-magnitude diagram (CMD) for the central region of NGC\,1313 using the F555W and F814W filters (Fig.~\ref{cmd}). The main sequence and the red supergiant plumes are well defined, while the blue supergiant plume and asymptotic giant branch are present, but more difficult to distinguish. We plotted stellar evolutionary tracks for ACS filters from the Padova group (L. Girardi 2006, private communication\footnote{\url{http://pleiadi.pd.astro.it/}}) to better understand the stellar content. To separate the massive stars in clusters from those in the star field background, we focus on the bright main sequence stars bluer than F555W-F814W=$-$1.0\,mag, and created two groups. The first group contains the brightest main sequence stars (m$_{F555W}$$<$22.7) while the second group includes fainter stars (22.7$\leq$m$_{F555W}$$<$24.2; red region). We exclude stars fainter than m$_{F555W}$=24.2 in the second group to avoid large photometric uncertainties. Figure~\ref{map}a maps the stars from both groups, clearly showing the spiral arms and bar of NGC\,1313. Figure~\ref{map}b shows the brightest main sequence stars from the first group only, which are essentially found to live in the most recent and compact star-forming regions along the bar and spiral arms. They are also often associated with F555W nebular emission. Figure~\ref{map}c maps the fainter main sequence stars of the second group. The morphological features of NGC\,1313 are still recognizable, but are not as sharp, indicating that these stars are largely part of the star field background. While fainter than the stars of the first group, these diffusely distributed stars are still very bright massive main sequence stars. According to the stellar evolutionary tracks, they have initial masses of about 8\,\msun\ to 20\,\msun, which corresponds to late O-type and early B-type stars. These mass values are lower limits considering the small extinction correction applied to the images (\S2). Such stars usually live for about 5 to 25\,Myr. 

We also create a third group (Fig.\,\ref{map}d) with stars fainter than the second group (24.2$<$m$_{F555W}$$<$25.0 and F555W-F814W $<$-0.44) but still brighter than the detection limit indicated on the CMD. According to the Padova stellar evolutionary tracks, the stars of this group have initial masses between 5 and 15\,\msun, which corresponds roughly to the spectral types from B5 to B1. Despite larger uncertainties and limitations due to the detection limit of this third group, the map shows that fainter stars (and presumably less massive) are even more spread than the second group (Fig.\,\ref{map}c).
This observation strongly supports the work of \citet{lada03} where stars form in clusters to disperse later on.

\section{On the Origin of the Diffuse UV emission}

The B-type stars observed in the diffuse field are similar to the stellar
population observed in inter-cluster regions by \citet{hoopes01} and \citet{chandar03,chandar05} in normal spiral and starburst galaxies.
There is no doubt that these B-type stars must contribute, at least in part, to the diffuse UV emission. Here we calculate the UV flux coming from these stars to see how it compares to the fraction of UV flux observed by \citet{hoopes01} and \citet{meurer95} in and outside clusters. Since the calculation is
based on optical colors (there is no information on individual massive stars in NGC\,1313 at shorter wavelengths), we do not expect to find strongly reliable quantitative values. However, we 
will have a good idea about the fraction of UV emission that stars alone can produce.  

First we extrapolate the UV flux produced by each star observed in the ACS images. To do this, we interpolate between Padova evolutionary tracks to estimate plausible values of the effective temperature and surface gravity for all stars on the CMDs (based on the three filters). Using these parameters, we assign a stellar atmosphere model from \citet{kurucz92} to each star and estimate the UV flux produced at 2200\AA. We then identify star clusters by eye and calculate the UV flux produced in and out of clusters. This was done by blurring the F435W image and identifying big and bright structures. The smaller clusters were lost in the image degradation. To quantify the errors of by eye cluster selection, we vary the numbers of clusters (from 2 to 13) and cluster sizes. We find that 90 to 96\% of UV emission at 2200\AA\ is produced outside the clusters for the central field of NGC\,1313. We also found that a larger homogenous extinction affects these fractions by less than 1\%.

We also calculate the flux around 4300\AA\ using the same method and compare it with the flux directly measured in the F435W image. We find that we underestimate the fraction of flux in clusters by no more than 15\%. Taking into account this uncertainty and those related to the number and size of clusters used for the calculations, we obtain that the stellar UV flux produced in and out of clusters is 10$^{+15}_{-6}$ and 90$^{+6}_{-15}$ percent, respectively. For starburst galaxies, we can expect the fraction of UV emission to be lower outside the clusters due to the larger number of recently born clusters in starburst galaxies compared to more normal spiral galaxies. These numbers are consistent with the works of \citet{meurer95}, \citet{hoopes01}, and \citet{chandar03,chandar05}.

We want to emphasize that the large majority of stars in the diffuse field of Figure~\ref{cmd} are not in
SOBAs or small clusters having a few OB stars, and that these objects can now be excluded as the principal source of the B-type stars in NGC\,1313. This can be seen by looking at the F555W image where there is no or very few nebular emission associated with small groups of OB stars. However, as discussed previously, we cannot tell for the most compact low-mass clusters (\S\ref{data}). Nevertheless, it is clear that massive stars are contributing to the diffuse UV emission, that they are single or in a very small cluster. Also, B-type stars in the field are sufficient to account for the diffuse UV emission in NGC\,1313, so there is no need to invoke scattering of photons by dust to explain the high fraction of UV emission coming from a diffuse background, though we cannot exclude this process entirely. It is important to note that scattered photons alone cannot explain why there are so many B-type stars in the field of NGC\,1313. Therefore, the scattered photons and small clusters (those with at least a few OB stars) scenarios must be rejected as the main source of diffuse UV emission, at least in NGC\,1313. HST/ACS data show that massive B-type stars {\it {not}} in clusters can account for the dominant diffuse UV emission in NGC\,1313.

\section{Evidence for Infant Mortality}
\label{infant}

\begin{figure}
\plotone{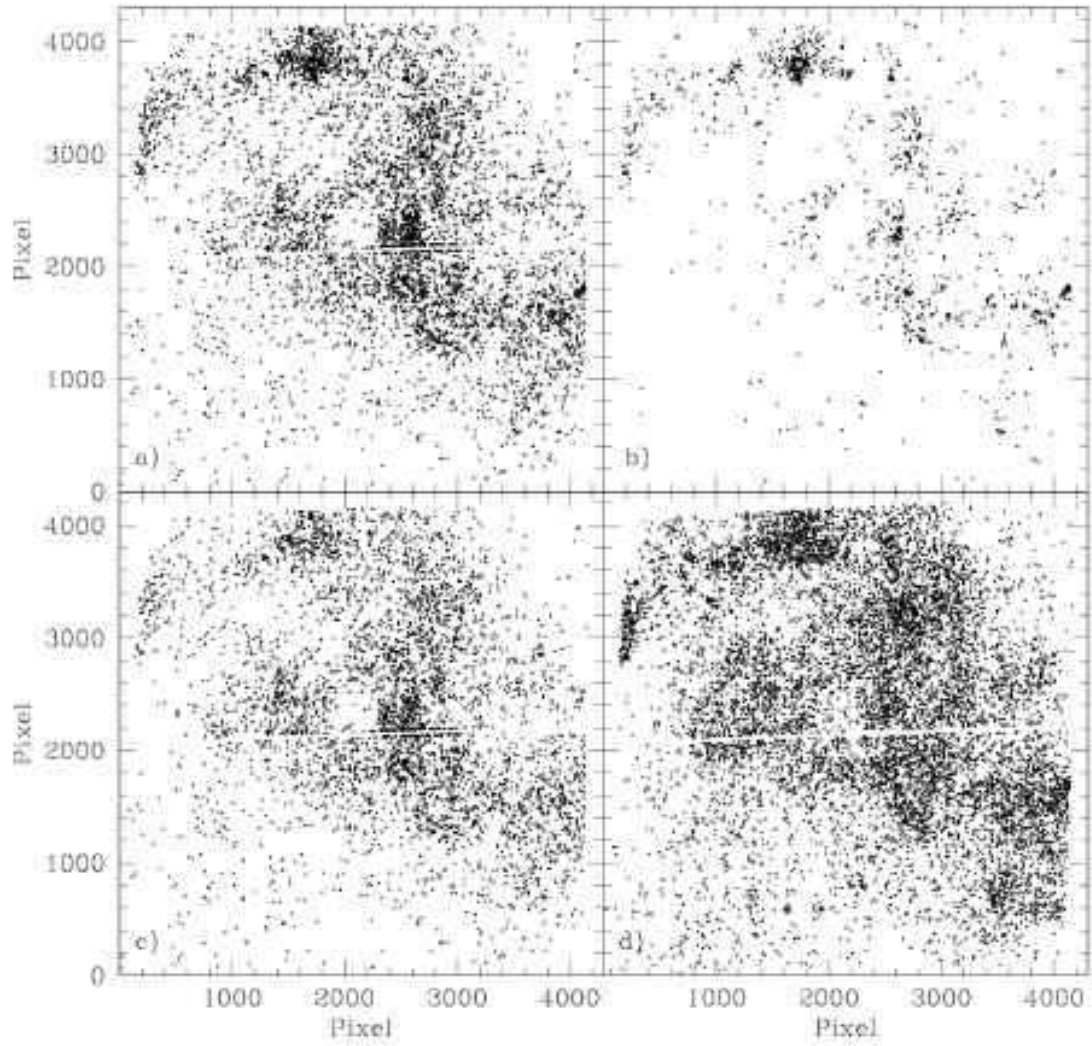}
\caption{\label{map}Spatial maps of main sequence stars selected from the CMD shown in Fig.~\ref{cmd}. a) Main sequence stars of the first two groups. b) Brightest stars located on the upper region of the CMD (first group). c)  Early B-type and late O-type stars within the lower region of the CMD (second group). d) Faintest stars located below the regions of the CMD brighter than m$_{F555W}$=25.0. See text for details.}
\end{figure}

Because most stars form in clusters \citep{lada03}, the fact that we find a large number of B-type stars in the field of NGC\,1313 means that the clusters in which they were born must have dissolved very rapidly, i.e. in less than their typical lifetime of 25\,Myr. If correcting the images by a modest internal extinction of E(B$-$V)$_{int}$=0.15, this would lead to a shorter dissolution timescale $\leq$10\,Myr. The dissolution processes must act very early in the life of the clusters and be very efficient. A phenomenon that can explain such a rapid dissolution is the infant mortality of the clusters (see \S\ref{intro}). Infant mortality has been invoke to explain the low number of clusters older than $\sim$10$^7$\,yr in the Antennae Galaxy and the SMC as discussed by \citet{fall04,chandar05,fall05}, and \citet{chandar06}. A dissolution timescale of 25\,Myr is a conservative
estimation since we corrected the images for a very modest amount of extinction. If the stars mapped in Fig.~\ref{map}c are more extinguished, they would then be more massive and younger than estimated here, leading to a smaller timescale. Also, it is not clear yet if the point sources detected in the field of NGC\,1313 are single stars of compact low-mass clusters because of the spatial resolution (see \S\ref{data}). If a significant fraction of point sources turned out to be clusters, we would then need to explain why there is a significant ongoing star formation in the field.

Several studies on cluster infant mortality are based on actively star-forming galaxies like the Antennae and the SMC. The large number of young stellar clusters ($<$10$^7$\,yr) in such systems might bias the age distribution function derived for the galaxies. The case of NGC\,1313 brings a different point of view in two ways. First, despite its relatively high star formation rate, NGC\,1313 is not a major starburst system and star formation presumably occurs less intensively during the last few 10$^8$\,Myr \citep{lar07}, which therefore avoids the biases potentially created by sudden burst of star formation. Second, we can examine the galaxy's stellar population directly, both in clusters and in the diffuse field. 

In this work, HST/ACS archival images reveal the presence of a large number of early B-type stars well spread in the star field background of NGC\,1313. The images show that there are not a large number of small star clusters or SOBAs, at least not enough to be the main source of diffuse UV emission in NGC\,1313. These stars are consistent with the B-type dominated stellar population observed by \citet{chandar03,chandar05} in inter-cluster environment of starburst galaxies. We calculated the UV flux at 2200\AA\ produced by these stars, showing that 90$^{+6}_{-15}$ percent of the flux in NGC\,1313 is produced outside of the clusters. Photon scattering is not required to explain the large fraction of UV emission outside clusters. Because of the short lifetime of the early B-type stars, we think that the infant mortality of star clusters is the only scenario left that can explain both the large number and the relatively homogeneity of the early B-type stars observed in NGC\,1313. We also propose that infant mortality of star clusters is the physical explanation for the diffuse UV emission observed in normal spiral galaxies, and probably in starburst galaxies.

\acknowledgments

This work was supported by HST grant HST-AR-10968.02-A and NASA Long-Term Space Astrophysics grant NAG5-9173.


\clearpage

\end{document}